\documentclass[prl,amsmath,twocolumn,floatfix,showpacs]{revtex4}
\usepackage{times}
\usepackage{graphicx}
\usepackage{amsmath}
\usepackage{mathptmx}

\newcommand{\parti}[2]{\frac{\partial #1}{\partial #2}}
\newcommand{\partit}[2]{\frac{\partial^2 #1}{\partial #2^2}}

\newcommand{\ket}[1]{|#1\rangle}

\newcommand{\avg}[1]{\left\langle #1 \right\rangle}
\newcommand{\update}[1]{#1}
\begin{document}

\title{Quantum Temporal Correlations and Entanglement via Adiabatic
Control of Vector Solitons}
\author{Mankei Tsang}
\email{mankei@optics.caltech.edu}
\date{\today}
\affiliation{
Department of Electrical Engineering, 
California Institute of Technology, Pasadena, CA 91125}
\begin{abstract}
It is shown that optical pulses with an average position accuracy beyond
the standard quantum limit can be produced by adiabatically expanding
an optical vector soliton followed by classical dispersion management.
The proposed scheme is also capable of entangling positions of optical
pulses and can potentially be used for general continuous-variable
quantum-information processing.
\end{abstract}

\maketitle
If an optical pulse consists of $N$ independent photons, then the
uncertainty in the pulse-center position is the pulse width divided by
$\sqrt{N}$, the so-called standard quantum limit
\cite{giovannetti_science}.  The ultimate limit permissible by quantum
mechanics, however, is determined by the Heisenberg uncertainty
principle and is smaller than the standard quantum limit by another
factor of $\sqrt{N}$, resulting in a quantum-enhanced accuracy useful
for positioning and clock synchronization applications
\cite{giovannetti_nature}.  To do better than the standard quantum
limit, a multiphoton state with positive frequency correlations and,
equivalently, negative time correlations is needed
\cite{giovannetti_nature}. Consequently, significant theoretical
\cite{coincident,tsang} and experimental \cite{kuzucu} efforts have
been made to create such a nonclassical multiphoton state. All
previous efforts were based on the phenomenon of spontaneous photon
pair generation in parametric processes, limiting $N$ to 2 only. The
resultant enhancement can only be regarded as a proof of concept and
is too small to be useful, considering that a large number of
uncorrelated photons can easily be obtained, with a standard quantum
limit orders of magnitude lower than the ultimate limit achievable by
two photons. It is hence much more desirable in practice to be able to
enhance the position accuracy of a large number of photons.  In this
Letter, for the first time to the author's knowledge, a scheme that
produces a multiphoton state with positive frequency correlations
among an arbitrary number of photons is proposed, thus enabling
quantum position accuracy enhancement for macroscopic pulses as
well. The scheme set forth therefore represents a major step forward
towards the use of quantum enhancement in future positioning and clock
synchronization applications.

The proposed scheme exploits the quantum properties of a vector
soliton, in which photons in different optical modes are bound
together by the combined effects of group-velocity dispersion,
self-phase modulation, and cross-phase modulation \cite{agrawal}. A
quantum analysis shows that the average position of the photons in a
vector soliton is insensitive to the optical nonlinearities and only
subject to quantum dispersive spreading, while the separations among
the photons is controlled by the balance between dispersion and
nonlinearities. These properties are in fact very similar to those of
scalar solitons \cite{lai,kartner}, so the idea of adiabatically
compressing scalar solitons for momentum squeezing \cite{fini} can be
similarly applied to vector solitons. To produce negative time
correlations, however, adiabatic soliton expansion should be performed
instead.  Given the past success of experiments on scalar quantum
solitons \cite{soliton_experiments} and vector solitons
\cite{vs_experiments}, the scheme set forth should be realizable with
current technology. The formalism should apply to spatial vector
solitons as well, so that the position accuracy of an optical beam can
be enhanced \cite{barnett}.  Moreover, the proposed scheme is capable
of creating temporal Einstein-Podolsky-Rosen (EPR) entanglement
\cite{epr} among the pulses in a vector soliton, suggesting that the
vector soliton effect, together with quantum temporal imaging
techniques \cite{tsang}, may be used for general continuous-variable
quantum-information processing \cite{braunstein}.

For simplicity, only vector solitons with two optical modes, such as
optical fiber solitons with two polarizations, are analyzed
in this Letter, although the results 
can be naturally extended to multimode vector
solitons, such as those studied in Refs.~\cite{crosignani}.
Two-mode vector solitons are classically described by the
coupled nonlinear Schr\"{o}dinger equations \cite{agrawal},
$i\parti{U}{t} = -b\partit{U}{z}+
2c(|U|^2  + B|V|^2)U$ and
$i\parti{V}{t} = -b\partit{V}{z}+
2c(|V|^2 + B|U|^2)V$,
where $U$ and $V$ are complex envelopes of the two polarizations,
assumed to have identical group velocities and group-velocity
dispersion, $t$ is the propagation time, $z$ is the longitudinal
position coordinate in the moving frame of the pulses, $b$ is the
group-velocity dispersion coefficient, $c$ is the self-phase
modulation coefficient, and $Bc$ is the cross-phase modulation
coefficient. For example, $B = 2/3$ for linear polarizations in a
linearly birefringent fiber \cite{menyuk}, $B = 2$ for circular
polarizations in an isotropic fiber \cite{berkhoer}, and $B = 1$
describes Manakov solitons \cite{manakov}, realizable in an
elliptically birefringent fiber \cite{menyuk}. $bc < 0$ is required
for solitons to exist.
The coupled nonlinear Schr\"odinger equations can be
quantized using the Hamiltonian
$\hat{H} = \hbar\int dz\mbox{ }[
b(\parti{\hat{U}^\dagger}{z}\parti{\hat{U}}{z} + 
\parti{\hat{V}^\dagger}{z}\parti{\hat{V}}{z}) + 
c(\hat{U}^\dagger\hat{U}^\dagger\hat{U}\hat{U}+
\hat{V}^\dagger\hat{V}^\dagger\hat{V}\hat{V}+
2B\hat{U}^\dagger\hat{V}^\dagger\hat{U}\hat{V})]$,
where $\hat{U}$ and $\hat{V}$ are photon annihilation operators of the
two polarizations and the daggers denote the corresponding creation
operators. The Heisenberg equations of motion derived from
this Hamiltonian are analyzed using perturbative techniques by
Rand \textit{et al.}\ \cite{rand}, who study the specific case of
Manakov solitons, and by Lantz \textit{et al.}\ \cite{lantz} and Lee
\textit{et al.}\ \cite{lee}, who numerically investigate the photon
number entanglement in higher-order vector solitons. As opposed to
these previous studies, in this Letter the exact quantum vector
soliton solution is derived in the Schr\"{o}dinger picture, in the
spirit of the scalar soliton analyses in Refs.~\cite{lai,kartner}.

Since the Hamiltonian conserves photon number in each mode and
the average momentum, one can construct simultaneous Fock and momentum
eigenstates with the Bethe ansatz
$\ket{n,m,p} = \frac{1}{\sqrt{n!m!}}\int d^n x\mbox{ } d^my\mbox{ }
f_{nmp}(x_1, ..., x_n, y_1, ..., y_m)\times
\hat{U}^\dagger(x_1) ... \hat{U}^\dagger(x_n)
\hat{V}^\dagger(y_1) ... \hat{V}^\dagger(y_m) \ket{0}$
\cite{lai,thacker},
where $n$ and $m$ are the photon numbers in the two polarizations and
$p$ is the average momentum.  Using the Schr\"{o}dinger equation
$E\ket{\Psi} =\hat{H}\ket{\Psi}$, one obtains
\begin{widetext}
\begin{align}
E_{nmp}f_{nmp}(x_1,...,x_n,y_1,...,y_m)
 &= \hbar\bigg\{
-b\sum_j\partit{}{x_j}-b\sum_k\partit{}{y_k}+
2c\Big[\sum_{i<j}\delta(x_j-x_i)+\sum_{l<k}\delta(y_k-y_l)+
B\sum_{j,k}\delta(x_j-y_k)\Big]\bigg\}
\times\nonumber\\&\quad
f_{nmp}(x_1,...,x_n,y_1,...,y_m).
\label{master}
\end{align}
The soliton solution of Eq.~(\ref{master}) is 
\begin{align}
f_{nmp}&= C_{nm}
\exp\bigg[ip\Big(\sum_j x_j + \sum_k y_k\Big)+
\frac{c}{2b}\Big(\sum_{i<j}|x_j-x_i|+\sum_{l<k}|y_k-y_l|+
B\sum_{j,k}|x_j-y_k|\Big)\bigg],\label{solution}
\end{align}
where $C_{nm}$ is a normalization constant.
The energy can be calculated by substituting Eq.~(\ref{solution})
into Eq.~(\ref{master}) and is given by
$E_{nmp}= \hbar bNp^2 -
\frac{\hbar c^2}{12b}[n(n^2-1)+m(m^2-1)+3B^2nm(n+m)]$,
where $N = n+m$.
A physical state should contain a distribution of momentum states,
say, a Gaussian, such that the time-dependent multiphoton
probability amplitude is now given by
\begin{align}
f_{nm} &= \int dp
\mbox{ } \frac{1}{(2\pi\Delta p^2)^{\frac{1}{4}}}
\exp\Big(-\frac{p^2}{4\Delta p^2}-ibNp^2t\Big)f_{nmp}
\label{pdistrib}\\
&=C_{nm}(8\pi)^{\frac{1}{4}}
\Big(\frac{\Delta p}{1+4ibN\Delta p^2 t}\Big)^{\frac{1}{2}}
\exp\bigg[-\frac{\Delta p^2}{1+4ibN\Delta p^2 t}
\Big(\sum_j x_j + \sum_k y_k\Big)^2\bigg]\times
\label{pulsecenter}
\\&\quad
\exp\bigg[\frac{c}{2b}\Big(\sum_{i<j}|x_j-x_i|+\sum_{l<k}|y_k-y_l|+
B\sum_{j,k}|x_j-y_k|\Big)\bigg],
\label{boundstate}
\end{align}
\end{widetext}
where $\Delta p$ is determined by initial conditions and a
constant energy term that does not affect the position
and momentum properties of a Fock state is omitted.  Although a
more realistic soliton state should have a superposition of Fock
states resembling a coherent state \cite{lai}, the Fock
components of a coherent state for $N >> 1$ have photon numbers very
close to the mean value, so a Fock state should be able to adequately
represent the position and momentum properties of a coherent-state
soliton.

The multiphoton amplitude $f_{nm}$ consists of two components; a
dispersive pulse-center component given by Eq.~(\ref{pulsecenter})
that governs the quantum dispersion of the average photon position
$\frac{1}{N}(\sum_j x_j+\sum_k y_k)$, and a bound-state component
given by Eq.~(\ref{boundstate}) that fixes the distances among the
photons via the attractive Kerr potentials. It follows
that the momentum-space probability amplitude, defined as the
$N$-dimensional Fourier transform of $f_{nm}$, also consists of an
average momentum component and a bound-state component that governs
the relative momenta among the photons.

If one increases $b$ or reduces $c$ adiabatically, the multiphoton
amplitude would remain in the same form, but with increased
uncertainties in the relative distances as well as reduced
uncertainties in the relative momenta. More crucially, the average
momentum uncertainty remains unaffected, leading to a multiphoton
state with positive momentum correlations.  The adiabatic
approximation remains valid if the change happens over a propagation
time scale $T >> \hbar/|E(t=T)-E(t=0)|$, which is on the order of the
initial soliton period divided by $N$.  As optical fiber solitons can
typically propagate for a few soliton periods before loss becomes a
factor, the desired adiabatic expansion should be realizable with
current technology. In the following it is assumed for simplicity that
only $c$ is adiabatically varied.  Mathematically, in the limit of
vanishing $c$, the bound-state component becomes relatively flat, and
$f_{nm}$ becomes solely governed by the pulse-center component,
\begin{align}
f_{nm} \propto \exp\bigg[-\frac{\Delta p^2}{1+4ibN\Delta p^2 t}
\Big(\sum_j x_j + \sum_k y_k\Big)^2\bigg].
\label{coincident1}
\end{align}
In the momentum space, as the bandwidth of the relative momenta
is reduced and becomes much smaller than the bandwidth of the average
momentum, the wavefunction in terms of momentum eigenstates becomes
\begin{align}
\ket{\Psi}&\propto \int dp\mbox{ }
\exp\Big(-\frac{p^2}{4\Delta p^2}-ibNp^2 t\Big)
\ket{n_p,m_p},
\label{coincident2}
\end{align}
where $\ket{n_p, m_p}$ denotes a momentum eigenstate with momentum $p$
and $n$ and $m$ photons in the respective polarizations. Except for
the dispersive phase term, Eq.~(\ref{coincident2}) is precisely the
desired coincident-frequency state that can achieve the ultimate limit
of average position accuracy \cite{giovannetti_nature}, as frequency is
trivially related to momentum via the dispersion relation. \update{ If
the pulse is sent across one channel only, adiabatic control of a
scalar soliton would already suffice for the purpose of temporal
uncertainty reduction, but the use of a vector soliton allows
quantum-enhanced pulses to be sent across different channels, as
originally envisioned by Giovannetti \textit{et al.}
\cite{giovannetti_nature}, for additional security.} The same
operation of position squeezing on a scalar soliton is previously
considered by Fini and Hagelstein, who nonetheless dismiss this
possibility due to the detrimental effect of quantum dispersion
\cite{fini}.

\begin{figure}[htbp]
\centerline{\includegraphics[width=0.45\textwidth]{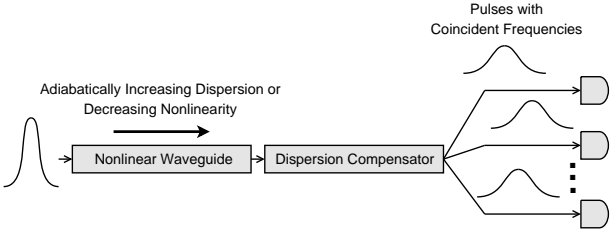}}
\caption{Proposed setup of generating multiphoton states with
quantum-enhanced average position accuracy.
}
\label{qvssetup}
\end{figure}

Fortunately, quantum dispersion, like classical dispersion, can be
compensated with classical dispersion management. If the vector
soliton propagates in another linear waveguide with an \emph{opposite}
group-velocity dispersion $b'$, \update{such that $b t = -b't'$, where
$t$ is the propagation time in the first waveguide and $t'$ is the
propagation time in the second waveguide,} then the dispersive phase
term in Eq.~(\ref{pdistrib}), $-ibNp^2 t$, can be cancelled, thus
restoring the minimum uncertainty in the average photon position,
while the pulse bandwidth remains constant because the second
waveguide is linear. The complete proposed setup is sketched in
Fig.~\ref{qvssetup}. To apply the scheme set forth to a spatial vector
soliton, negative refraction \cite{pendry} is required to compensate
for the quantum diffraction instead.

In order to understand how the quantum vector soliton solution
corresponds to a classical soliton in typical experiments,
consider the specific case of a Manakov soliton, where $B =
1$. Other vector solitons should have very similar properties given
the similarity of the solutions. If the photon position variables are
re-indexed in the following new notations $\{z_1,...,z_{N}\} =
\{x_1,...,x_n,y_1,...,y_m\}$, the multiphoton amplitude in
Eqs.~(\ref{pulsecenter}) and (\ref{boundstate}) becomes
\begin{align}
f_{nm} &= C_{nm}(8\pi)^{1/4}
\Big(\frac{\Delta p}{1+4ibN\Delta p^2 t}\Big)^{\frac{1}{2}}
\times\nonumber\\&\quad
\exp\bigg[-\frac{\Delta p^2}{1+4ibN\Delta p^2t}
\Big(\sum_j z_j\Big)^2+
\frac{c}{2b}\sum_{i<j}|z_j-z_i|\bigg].
\label{laplacian}
\end{align}
Intriguingly, this solution is exactly the same as the scalar soliton
solution \cite{lai}, or in other words, a Manakov soliton is
quantum-mechanically equivalent to a scalar soliton. This equivalence
explains the discovery by Rand \textit{et al.}\ that the squeezing
effect of a Manakov soliton has the same optimum as a scalar soliton
\cite{rand}. Moreover, $C_{nm}$ can now be borrowed from the scalar
soliton analysis and is given by $C_{nm} =
[(N-1)!|c/b|^{N-1}/(2\pi)]^{1/2}$ \cite{lai}.  The knowledge of
$C_{nm}$ allows one to calculate the correlations among the photon
positions using standard statistical mechanics techniques.  An
expression for $\avg{\sum_{i<j}|z_j-z_i|}$ can be derived, and, by
symmetry,
\begin{align}
\avg{|z_j-z_i|} &= \frac{1}{N(N-1)}\avg{\sum_{i<j}|z_j-z_i|}
=\Big|\frac{2b}{Nc}\Big| \sim W_0.
\end{align}
As expected, the mean absolute distance between any two photons is
on the order of the classical soliton pulse width, $W_0\sim |2b/(Nc)|$
\cite{lai}.  Next, assume that the variance of the relative distance
is related to the square of the mean absolute distance by a
parameter $q$,
\begin{align}
\avg{|z_j-z_i|^2} &= q\avg{|z_j-z_i|}^2 = \frac{4qb^2}{N^2c^2}.
\label{alpha}
\end{align}
While an explicit expression for $q$ is hard to derive, $q$ must
depend only on $N$ by dimensional analysis, must be larger than 1
because $\avg{|z_j-z_i|^2}\ge\avg{|z_j-z_i|}^2$, and is likely to be
on the order of unity, as will be shown later.  By symmetry,
$\avg{|z_j-z_i|^2} = \avg{z_j^2}-2\avg{z_iz_j}+\avg{z_i^2} =
2\avg{z_j^2}-2\avg{z_iz_j}$. Eq.~(\ref{alpha}) then gives
\begin{align}
\avg{z_j^2}-\avg{z_iz_j} &= \frac{2qb^2}{N^2c^2}.
\label{l2norm}
\end{align}
Furthermore, the variance of $\sum_j z_j$ is simply given by
\begin{align}
\avg{\Big(\sum_j z_j\Big)^2} &= 
N\avg{z_j^2}+N(N-1)\avg{z_iz_j}
=\Delta z^2, \label{varsum2}
\end{align}
where
$\Delta z^2 = 1/(4\Delta p^2) + 4(bN \Delta p t)^2$.
From Eqs.~(\ref{l2norm}) and (\ref{varsum2}) the
covariances can be obtained explicitly,
\begin{align}
\avg{z_j^2} &= \frac{\Delta z^2}{N^2}+\frac{2q(N-1)b^2}{N^3c^2},
\quad
\avg{z_iz_j} = \frac{\Delta z^2}{N^2}-\frac{2qb^2}{N^3c^2}.
\label{covar}
\end{align}
A quantum soliton solution best resembles a classical initial
condition with independent photons when the initial covariance
is zero,
$\avg{z_iz_j}_{t=0} = 1/(4N^2\Delta p^2) -
2qb^2/(N^3 c^2)|_{t=0} = 0$, and
\begin{align}
\Delta p &= \Big|\frac{\sqrt{N}c}{\sqrt{8q}b}\Big|_{t=0}
\sim \frac{1}{2\sqrt{N}W_0}.
\label{sqlmomentum}
\end{align}
Incidentally, the average momentum uncertainty $\Delta p$ is at the
shot-noise level when the photons are initially uncorrelated. This
justifies the assumption that $q$ is on the order of unity.  An
initial condition with independent photons would then mostly couple to
a soliton state with $\Delta p$ given by Eq.~(\ref{sqlmomentum}),
while coupling to continuum states should be negligible. Adiabatically
increasing $|b/c|$ thus makes $\avg{z_iz_j}$ negative and therefore
introduces the necessary negative correlations among the photon
positions.  To investigate the magnitude of the quantum enhancement in
practice, it is useful to compare the quantum theory to the classical
soliton theory, as the two regimes should converge when $N >> 1$ and
$\gamma << \sqrt{N}$. According to the classical theory, if the ratio
between the final and initial values of $|b/c|$ is $\gamma$, the pulse
bandwidth is also reduced by a factor of $\gamma$, from $\sim 1/W_0$
to $\sim 1/(\gamma W_0)$.  Since the final average position uncertainty
is the same as the input value, the accuracy enhancement over the standard
quantum limit, for the same reduced bandwidth $\sim 1/(\gamma W_0)$,
is hence also given by $\gamma$, in the regime of moderate pulse
expansion $\gamma << \sqrt{N}$. Because the ultimate soliton state,
given by Eqs.~(\ref{coincident1}) and (\ref{coincident2}), has a
bandwidth given by Eq.~(\ref{sqlmomentum}), $\sim 1/(\sqrt{N}W_0)$,
the ultimate limit is reached only when $\gamma >> \sqrt{N}$.

As the photons across different optical modes become correlated via
the cross-phase modulation effect, entanglement is expected among the
pulse positions in a vector soliton. To estimate the magnitude of the
entanglement in terms of macroscopic position variables, consider
again the case of Manakov solitons. Let the pulse-center coordinates
of the respective polarizations be $X$ and $Y$, defined as $X =
\frac{1}{n}\sum_{j=1}^{n} z_j$ and $Y = \frac{1}{m}\sum_{k=m+1}^N z_k$.  
If $n = m = N/2$ is assumed for simplicity,
the following statistics for $X$ and $Y$ can be derived using
Eqs.~(\ref{covar}),
\begin{align}
\avg{\left(\frac{X+Y}{2}\right)^2} &= \frac{\Delta z^2}{N^2},
\quad
\avg{\left(\frac{X-Y}{2}\right)^2} = \frac{2qb^2}{N^3c^2}.
\label{posdiff}
\end{align}
Similar to a two-photon vector soliton \cite{tsang}, the average
position of the two pulses is affected by quantum dispersion, while
the relative distance is bounded by the Kerr effect. For two initially
uncorrelated pulses, the two expressions in Eq.~(\ref{posdiff}) have
the same value. If, however, $b$ and $c$ are adiabatically manipulated,
then the nonlocal uncertainty product
$\avg{(X-Y)^2}\avg{(P_X-P_Y)^2}$, where $P_X$ and $P_Y$ are the
conjugate momenta, can remain constant under the adiabatic
approximation, while $\avg{(X-Y)^2}$ and $\avg{(P_X-P_Y)^2}$ can be
arbitrarily varied. Since $\avg{(P_X+P_Y)^2}$ always remains constant
and $\avg{(X+Y)^2}$ can also remain the same as the input value if
quantum dispersion is compensated, $\avg{(X-Y)^2}\avg{(P_X+P_Y)^2}$ or
$\avg{(X+Y)^2}\avg{(P_X-P_Y)^2}$ can be arbitrarily reduced, thus
resulting in EPR entanglement. Combined with quantum temporal imaging
techniques, which are able to temporally reverse, compress, and expand
photons in each mode \cite{tsang}, adiabatic vector soliton control
potentially provides a powerful way of fiber-based continuous-variable
quantum-information processing \cite{braunstein}.

Discussions with Demetri Psaltis
and financial support by the Engineering Research Centers Program of
the National Science Foundation under Award Number EEC-9402726 and the
Defense Advanced Research Projects Agency (DARPA) are gratefully
acknowledged.

\pagebreak

\section{Erratum}
On closer inspection, it is found that the soliton solution given by
Eq.~(2) and the expression for the energy $E_{mnp}$ below Eq.~(2) in
the published Letter \cite{tsang_prl} are valid only for the Manakov
soliton ($B = 1$) or uncoupled scalar solitons ($B = 0$).  This can be
seen by observing that the energies calculated using Eqs.~(1) and (2)
in Ref.~\cite{tsang_prl} in different regions of the configuration space
are different unless $B = 1$ or $B = 0$.  For example, for $n = 2$ and
$m = 1$, the energy in the regions of $x_1<y_1<x_2$ and $x_2<y_1<x_1$
calculated using Eq.~(2) is different from the energy in the other
regions, unless $B = 1$ or $B = 0$.

This error does not affect the rest of Ref.~\cite{tsang_prl} that focuses
on the Manakov soliton. Furthermore, it can be argued on physical
grounds that the proposed scheme should still work for any positive
$B$, despite the lack of an explicit soliton solution in general.
First, it can be shown using the ``center-of-mass'' coordinate system
\cite{hagelstein} that all stable solutions of the Schr\"odginer
equation given by Eq.~(1) in Ref.~\cite{tsang_prl} must consist of a
dispersive center-of-mass component and a bound-state component.  For
any positive $B$, the bound-state component must become flat
in the ultimate limit of adiabatic pulse expansion, so any stable
solution still approaches the form of Eqs.~(6) and (7) in
Ref.~\cite{tsang_prl}. Second, the average timing jitter of the two pulses
is affected only by dispersion, regardless of $B$.  If the net
dispersion is zero and the fiber is assumed to be lossless, the output
jitter is the same as the input jitter.  As long as the solitary wave
propagation is stable against the adiabatic change in fiber
parameters, the bandwidth can be adiabatically reduced, leading to a
raised timing standard quantum limit.  Hence the final jitter must be
lower than the raised standard quantum limit regardless of $B$.
Finally, Eq.~(1) in Ref.~\cite{tsang_prl} shows that the cross-phase
modulation effect manifests itself as an attractive potential among
photons across the two polarizations. In a stable solution of Eq.~(1),
the relative position and relative momentum of the two pulses can
therefore be controlled by adjusting the balance between dispersion
and cross-phase modulation. Since temporal entanglement has been
proved to occur for the Manakov soliton in Ref.~\cite{tsang_prl}, this
physical picture should remain qualitatively the same for a different
cross-phase modulation coefficient, and temporal entanglement should
also occur for any positive $B$.

There are also two typographical errors in \cite{tsang_prl}. The factor
$\frac{1}{N(N-1)}$ in Eq.~(9) should read $\frac{2}{N(N-1)}$, and in
the second-to-last paragraph, $Y = (1/m)\sum_{k=m+1}^N z_k$ should
read $Y = (1/m)\sum_{k=n+1}^N z_k$.  These typographical errors do not
affect the rest of the Letter in any way.


\begin{thebibliography}{23}
\bibitem{giovannetti_science} V.\ Giovannetti, S.\ Lloyd, and L.\ Maccone,
Science \textbf{306}, 1330 (2004).

\bibitem{giovannetti_nature} V.\ Giovannetti,  S.\ Lloyd, and L.\ Maccone,
\nat \textbf{412}, 417 (2001).

\bibitem{coincident} V.\ Giovannetti \textit{et al.},
\prl \textbf{88}, 183602 (2002);
Z.\ D.\ Walton \textit{et al.},
\pra \textbf{67}, 053810 (2003);
J.\ P.\ Torres \textit{et al.},
\ol \textbf{30}, 314 (2005);
M.\ Tsang and D.\ Psaltis,
\pra \textbf{71}, 043806 (2005).

\bibitem{tsang} M.\ Tsang and D.\ Psaltis,
\pra \textbf{73}, 013822 (2006).

\bibitem{kuzucu} O.\ Kuzucu \textit{et al.},
\prl {\bf 94}, 083601 (2005).

\bibitem{agrawal} G.\ P.\ Agrawal,
\textit{Nonlinear Fiber Optics}
(Academic Press, San Diego, 2001).

\bibitem{lai} Y.\ Lai and H.\ A.\ Haus,
\pra \textbf{40}, 844 (1989);
\textit{ibid.} \textbf{40}, 854 (1989).

\bibitem{kartner} F.\ X.\ K\"{a}rtner and H.\ A.\ Haus,
\pra \textbf{48}, 2361 (1993);
P.\ L.\ Hagelstein,
\textit{ibid.}\ \textbf{54}, 2426 (1996).

\bibitem{fini} J.\ M.\ Fini and P.\ L.\ Hagelstein,
\pra \textbf{66}, 033818 (2002).

\bibitem{soliton_experiments} See, for example,
A.\ Sizmann,
Appl.\ Phys.\ B \textbf{65}, 745 (1997), and references therein;
Ch.\ Silberhorn \textit{et al.},
\prl \textbf{86}, 4267 (2001).

\bibitem{vs_experiments} See, for example,
M.\ N.\ Islam,
\ol \textbf{14}, 1257 (1989);
J.\ U.\ Kang \textit{et al.},
\prl \textbf{76}, 3699 (1996);
Y.\ Barad and Y.\ Silberberg,
\textit{ibid.}\ \textbf{78}, 3290 (1997);
S.\ T.\ Cundiff \textit{et al.},
\textit{ibid.}\ \textbf{82}, 3988 (1999).

\bibitem{barnett} S.\ M.\ Barnett, C.\ Fabre, and A.\ M\^{a}itre,
Eur.\ Phys.\ J.\ D \textbf{22}, 513 (2003).

\bibitem{epr} A.\ Einstein, B.\ Podolsky, and N.\ Rosen,
Phys.\ Rev.\ \textbf{47}, 777 (1935).

\bibitem{braunstein} S.\ L.\ Braunstein and P.\ van Loock,
\rmp \textbf{77}, 513 (2005).

\bibitem{crosignani}B.\ Crosignani and P.\ Di Porto,
\ol \textbf{6}, 329 (1981);
F.\ T.\ Hioe,
\prl \textbf{82}, 1152 (1999).

\bibitem{menyuk} C.\ R.\ Menyuk,
\jqe \textbf{25}, 2674 (1989).

\bibitem{berkhoer} A.\ L.\ Berkhoer and V.\ E.\ Zakharov,
Sov.\ Phys.\ JETP \textbf{31}, 486 (1970).

\bibitem{manakov} S.\ V.\ Manakov,
Sov.\ Phys.\ JETP \textbf{38}, 248 (1974).

\bibitem{rand} D.\ Rand, K.\ Steiglitz, and P.\ R.\ Prucnal,
\pra \textbf{71}, 053805 (2005).

\bibitem{lantz} E.\ Lantz \textit{et al.},
J.\ Opt.\ B \textbf{6}, S295 (2004).

\bibitem{lee} R.-K.\ Lee, Y.\ Lai, and B.\ A.\ Malomed,
\pra \textbf{71}, 013816 (2005).

\bibitem{thacker}H.\ B.\ Thacker,
\rmp \textbf{53}, 253 (1981).

\bibitem{pendry} V.\ G.\ Veselago,
Sov.\ Phys.\ Usp.\ \textbf{10}, 509 (1968);
J.\ B.\ Pendry,
\prl \textbf{85}, 3966 (2000).

\end{thebibliography}

\begin{thebibliography}{2}
\bibitem{tsang_prl} M.\ Tsang, \prl \textbf{97}, 23902 (2006).
\bibitem{hagelstein}P.\ L.\ Hagelstein, \pra \textbf{54}, 2426 (1996).
\end{thebibliography}
\end{document}